\documentclass{aa}
\usepackage{graphicx}

\begin{document}

\title{Microscopic study of neutrino trapping in hyperon stars} 

\author{I.\ Vida\~na\inst{1} \and I. Bombaci\inst{1} \and A.\ Polls\inst{2} \and A.\ Ramos\inst{2}}
\institute{Dipartimento di Fisica, Universit\`a di Pisa 
and INFN Sezione di
Pisa,Via Buonarroti 2, I-56127 Pisa, Italy
\and 
Departament d'Estructura i Constituents de la Mat\`eria,
Universitat de Barcelona, E-08028 Barcelona, Spain}

\date{Received 4 September 2002/ Accepted 29 November 2002}

\abstract{
Employing the most recent parametrization of the baryon-baryon interaction of 
the Nijmegen group, we investigate, in the framework of the 
Brueckner--Bethe--Goldstone many-body theory at zero temperature, 
the influence of neutrino trapping on the composition, equation of state, and 
structure of neutron stars, relevant to describe the physical conditions of 
a neutron star immediately after birth (protoneutron star). 
We find that the presence of neutrinos changes significantly the 
composition of matter delaying the appearance of hyperons and making the 
equation of state stiffer. We explore the consequences of neutrino trapping 
on the early evolution of a neutron star and on the nature of the final compact 
remnant left by the supernova explosion.  

\\
}

\maketitle


\section{Introduction}
Neutrinos play a crucial role in the physics of supernova explosions 
(Janka and M\"uller 1996) and in the early evolution of their compact stellar 
remnants (Burrows and Lattimer 1986, Janka and M\"uller 1995). 
During the collapse of the pre-supernova core, a large number of neutrinos is 
produced by electron capture process. 
Immediately following the core bounce the radius of the newly formed neutron 
star shrinks from about 100~km to about 10~km. 
During this same period (up to about 1~second after core bounce) substantial 
matter accretion occurs on the compact star (this accretion may eventualy 
led to the formation of a black hole). 
As the newly formed neutron star contracts the neutrino mean free path 
$\lambda_\nu$ decreases, and above a critical value of the density 
({\it neutrino trapping density}) $\lambda_\nu$ becomes smaller than the 
stellar radius. 
Under these physical conditions  neutrinos are {\it trapped} in the star, 
{\it i.e.,} the neutrino diffusion time is of the order of a few tens of seconds. 
Neutrino trapping has a strong influence on the overal {\it stiffness} of the 
equation of state (EoS) of dense stellar matter. 
Thus, the physical conditions of the hot and lepton-rich newborn 
neutron star (the so-called protoneutron star) differ substantially from 
those of the cold and deleptonized neutron star.  
Nevertheless, this stage nearly fulfills the conditions of hydrostatical 
equilibrium (Burrows \& Lattimer 1986).
 
The composition and the structure of protoneutron stars have been systematically 
investigated by Prakash {\it et al.} (1997) and by Strobel {\it et al.} (1999) using a 
large sample of modern equations of state of dense stellar matter. 
The implications of the early evolution of a protoneutron star on the concept 
of neutron star maximum mass have been studied by the authors of Refs. 
(Bombaci 1996, Prakash {\it et al.} 1997, Strobel \& Weigel 2001).        
 
Due to the rapid increase of the nucleon chemical potentials with density, 
hyperons ($\Lambda$, $\Sigma^{-}$, $\Sigma^{0}$, $\Sigma^{+}$, $\Xi^{-}$ 
and $\Xi^{0}$ particles) are expected to appear in the core of neutron stars, 
as suggested in the pioneer work by Ambartsumyan and Saakyan (1960). 
Since then the structrural properties  of these {\it hyperon stars}  
have been studied by many researchers using a variety of aproaches (see {\it e.g.,} Pandharipande (1971), 
Glendenning (1985), Keil and Janka 1994, Shaffner and Mishustin (1996), 
Prakash {\it et al.} (1997), Balberg and Gal (1997), Baldo {\it et al.} (2000), 
Vida\~na {\it et al.} (2000a)). 

All the previous studies of 
hyperonic matter with trapped neutrinos have been done in the 
framework of a relativistic theoretical field model of nucleons and hyperons 
interacting via meson exchange in a mean field approximation (Keil and Janka 1994, 
Prakash {\it et al.} 1997).     
In the present work, we use a microscopic approach instead, 
which is based on the Brueckner--Bethe--Goldstone (BBG) many body theory. 
In our calculations the basic input is the baryon-baryon interaction 
for the complete baryon octet  ($n$, $p$, $\Lambda$, $\Sigma^{-}$, $\Sigma^{0}$, 
$\Sigma^{+}$, $\Xi^{-}$ and $\Xi^{0}$)     
developed recently by Stoks and Rijken (1999).    
Within this approach we compute the EoS of hyperonic matter with trapped 
neutrinos and the corresponding properties of newborn hyperon stars.  
A similar microscopic approach has been recently employed by Baldo {\it et al.} (2000) 
and Vida\~na {\it et al.} (2000a) to study cold and deleptonized hyperon stars. The primery 
purpose of the present work is to investigate the effects of neutrino trapping on the structure
and evolution of newly formed hyperon stars.  

The paper is organized in the following way. A brief review of the Brueckner--Hartree--Fock (BHF) approximation of the BBG many-body theory at zero 
temperature extended to the hyperonic sector is given in Sec. \ref{sec:sec2.1}. Equilibrium conditions and Eos of $\beta$-stable matter are 
discussed in Sec. \ref{sec:sec2.2}. Section \ref{sec:sec3} is devoted to the presentation and discussion of the results. Finally, a short summary 
and the main conclusions of this work are drawn in Sec. \ref{sec:sec4}.


\section{Equation of State and equilibrium conditions}
\label{sec:sec2}

\subsection{Many-body theory of hyperonic matter}  
\label{sec:sec2.1}

Our calculation of the EoS of high density matter is based on the 
BHF approximation of the BBG many-body theory 
at zero temperature extended to the hyperonic sector (Baldo {\it et al.} 2000, 
Vida\~na {\it et al.} 2000a). 
We start it by constructing all baryon-baryon (nucleon-nucleon (NN), 
hyperon-nucleon (YN) and hyperon-hyperon (YY)) $G$-matrices, which describe in an 
effective way the interactions between baryons in the presence of a surrounding 
hadronic medium. They are formally obtained by solving the well known 
Bethe--Goldstone equation, written schematically as
\begin{equation}
\begin{array}{l}
   G(\omega)_{B_1B_2,B_3B_4} =
   V_{B_1B_2,B_3B_4}+ \\
  \displaystyle{ \sum_{B_5B_6}V_{B_1B_2,B_5B_6}   
   \frac{Q_{B_5B_6}}{\omega-E_{B_5}-E_{B_6}+ i\eta}
   G(\omega)_{B_5B_6,B_3B_4} } \ .
\end{array}
   \label{eq:gmatrix}
\end{equation}

In the expression above the first (last) two subindices indicate the initial 
(final) two-baryon states compatible with a given value $S$ of the strangeness 
(NN for $S=0$, YN for $S=-1,-2$, and YY for $S=-2,-3,-4$), $V$ is the bare 
baryon-baryon interaction,  $Q$ is the Pauli operator which allows only intermediate 
two-body states compatible with the Pauli principle, and $\omega$ is the 
so-called starting energy.

The single-particle energy of a baryon $B_i$ is given by
(we use units in which $\hbar =1$, $c=1$)  
\begin{equation}
E_{B_i}=M_{B_i}+\frac{k^2}{2M_{B_i}}+U_{B_i}(k) \ ,
\label{eq:spe}
\end{equation}
where $M_{B_i}$ denotes the rest mass of the baryon,  
and the single-particle potential energy $U_{B_i}(k)$ represents the averaged 
field ``felt'' by the baryon due to its interaction with the other baryons of the 
medium. In the BHF approximation $U_{B_i}(k)$ is given by
\begin{equation}
\begin{array}{l}
       U_{B_i}(k) = \nonumber \\
       \displaystyle{
       {\mathrm Re} \sum_{B_j}\sum_{k' \leq k_{F_{B_j}}}
       \left\langle \vec{k}\vec{k'}\right |
       G_{B_iB_j,B_iB_j}(\omega=E_{B_i}+E_{B_j})
       \left | \vec{k}\vec{k'} \right\rangle } \ ,
\end{array}
\label{eq:upot}
\end{equation}
where a sum over all the Fermi seas of the different baryon species is performed, 
and the matrix elements are properly antisymmetrized when baryons $B_i$ and $B_j$ 
belong to the same isomultiplet. We note here that Brueckner-type calculations are very time
consuming since one has to solve a self-consistent set of coupled-channel equations for the different
strangeness sectors (see Vida\~na {\it et al.} 2000b for details). Therefore, in order to do the calculations  
less time consuming we have adopted the so-called discontinuous prescription for the single-particle energy ({\it i.e.,} 
$E_{B_i}=M_{B_i}+k^2/2M_{B_i}$ for $k > k_{F_{B_i}}$) when solving the Bethe--Goldstone equation.
The present calculations have been carried out by using the most recent 
parametrization of the bare baryon-baryon potential for the complete baryon octet 
as defined by Stoks and Rijken (1999).  
This potential model, which aims at describing all interaction channels with 
strangeness from $S=0$ to $S=-4$, is based on SU(3) extensions of the Nijmegen 
nucleon-nucleon and hyperon-nucleon potentials (Rijken {\it et al.} 1998). 

Once a self-consistent solution of Eqs. (\ref{eq:gmatrix})--({\ref{eq:upot}) is
achieved, the baryonic energy density $\varepsilon_b$ can be evaluated in the BHF 
approximation according to the following expression: 
\begin{equation}
\begin{array}{r}
\displaystyle{
\varepsilon_b = 2\sum_{B_i} \int_0^{k_{F_{B_i}}} \frac{d^3 k}{(2\pi)^3}
\left(M_{B_i} + \frac{k^2}{2M_{B_i}}+ \right. } \nonumber \\
\displaystyle{
\left.
\frac{1}{2}U_{B_i}^N(k)+\frac{1}{2}U_{B_i}^Y(k)
 \right) }\ ,
\end{array}
\label{eq:edb}
\end{equation}
where we have split, according to Eq.\ (\ref{eq:upot}) the baryon single-particle potential
$U_{B_i}$ into a contribution, $U_{B_i}^N$, coming from the interaction of the
baryon $B_i$ with  all the nucleons of the system, and a contribution, $U_{B_i}^Y$,
coming from the interaction with the hyperons.

It is well known that non-relativistic many-body calculations, based on 
purely two-body forces, fail to reproduce the empirical saturation point 
for symmetric nuclear matter and the binding energy and radius of light nuclei. 
The remedy to the previous deficiency is  to introduce three-body forces (TBF) 
between nucleons. 
In hyperonic matter the repulsion induced at high densities by nucleon three-body 
forces enhances substancially the hyperon population which in turn induces a 
strong softening of the EoS (Schulze {\it et al.} 1998, Baldo {\it et al.} 2000). 

In order to include the effects of TBF between nucleons in our computational scheme, 
we have replaced the pure nucleonic contribution to the baryonic energy density 
$\varepsilon_b$ (Eq.\ \ref{eq:edb}), {\it i.e.,}
\begin{equation}
\varepsilon_{NN} \equiv 2\sum_{N_i} \int_0^{k_{F_{N_i}}} \frac{d^3 k}{(2\pi)^3}
\left(M_{N_i} + \frac{k^2}{2M_{N_i}}+\frac{1}{2}U_{N_i}^N(k) \right) \ ,
\label{eq:enn}
\end{equation}
by the analytic parametrization developed 
by Heiselberg and Hjorth-Jensen (1999)
\begin{equation}
\varepsilon_{NN}=\rho_N\left(M_N+E_0u\frac{u-2-\delta}{1+u\delta}+S_0u^{\gamma}(1-2Y_p)^2\right) \ .
\label{eq:hh99}
\end{equation}
Here $u=\rho_N/\rho_0$ is the ratio of the nucleonic density to nuclear saturation density ($\rho_0=0.16$ fm$^{-3}$) and $Y_p=\rho_p/\rho_N$ 
is the proton fraction. This approach parametrizes the nucleon energy density obtained from the variational calculation using the Argonne 
$V_{18}$ nucleon-nucleon interaction with three-body forces and relativistic boost corrections of Akmal {\it et al.} (1998). The best 
fit of this simple functional is obtained for $E_0=15.8$ MeV, $S_0=32$ MeV, $\gamma=0.6$ and $\delta=0.2$ (see Ref. Heiselberg and Hjorth-Jensen 1999 
for more details). Therefore the baryonic energy density will be given by
\begin{equation}
\varepsilon_b=\varepsilon_{NN}+\varepsilon' \ ,
\label{eq:eb}
\end{equation}
with $\varepsilon_{NN}$ obtained from Eq.\ (\ref{eq:hh99}) and
\begin{equation}
\begin{array}{l}
\displaystyle{
\varepsilon' = 2\sum_{N_i} \int_0^{k_{F_{N_i}}} \frac{d^3 k}{(2\pi)^3}
\frac{1}{2}U_{N_i}^Y(k)+ } \nonumber \\
\displaystyle{
2\sum_{Y_i} \int_0^{k_{F_{Y_i}}} \frac{d^3 k}{(2\pi)^3}
\left( M_{Y_i}
+\frac{k^2}{2M_{Y_i}}+ \right. } \nonumber \\
\displaystyle{
\left.
\frac{1}{2}U_{Y_i}^N(k)+\frac{1}{2}U_{Y_i}^Y(k)
\right) } \ . 
\end{array}   
\label{eq:ebt}
\end{equation}


\subsection{Equilibrium conditions and EoS of $\beta$-stable matter}
\label{sec:sec2.2}

The concentrations of the different constituents in the stellar interior are 
determined by the requirements of electric charge neutrality and equilibrium under 
weak interaction processes (``chemical'' equilibrium)  
\begin{eqnarray}
B_1 \rightarrow B_2 + \ell + {\overline \nu}_\ell \,, \qquad 
B_2 + \ell \rightarrow B_1 + \nu_\ell
\label{eq:bproc}
\end{eqnarray}
where $B_1$ and $B_2$ are baryons, and $\ell$ is a lepton ($e^-$ or $\mu^-$) and 
$\nu_\ell$ (${\overline \nu}_\ell$) is the associated neutrino (antineutrino). 
For stellar matter with trapped neutrinos, these two requirements imply that 
the relations 
\begin{equation}
\sum_i \rho_{B{_i}}^{(+)} + \sum_{\ell} \rho_{\ell}^{(+)} = 
\sum_i \rho_{B{_i}}^{(-)} + \sum_{\ell} \rho_{\ell}^{(-)}
\label{eq:charge}  
\end{equation}
\begin{equation}
\mu_i = b_i\mu_n - q_i(\mu_\ell -\mu_{\nu_\ell}) \, ,
\label{eq:tbeta}
\end{equation}
are satisfied. Above, $\rho_{B{_i}}$ ($\rho_{\ell}$) denotes the baryon (lepton)  
number density and the superscripts $(\pm)$ on $\rho_{B{_i}}$ ($\rho_{\ell}$) 
signify positive or negative electric charge. 
The symbol $\mu_i$ refers to the chemical potential of baryon of 
the species $i$, $b_i$ is its baryon number, and $q_i$ is its charge. 
The chemical potential of the neutron is denoted by $\mu_n$, and 
the chemical potential of the neutrino $\nu_\ell$ is denoted by  $\mu_{\nu_\ell}$. 
Because neutrinos are trapped in the star, the lepton number per baryon $Y_{L\ell}$
of each lepton flavor must be conserved  on dynamical time scales 
\begin{eqnarray}
 Y_{Le} = Y_e + Y_{\nu_e} \,, \qquad  Y_{L\mu} = Y_\mu+ Y_{\nu_\mu} \, .
\label{eq:lepnum}
\end{eqnarray}
Gravitational collapse calculations of the core of massive stars indicate that 
at the onset of trapping the electron lepton fraction 
$Y_{Le}=Y_e+Y_{\nu_e} \approx 0.4$. In addition, as the trapping in supernova 
occurs when the collapsing core reaches densities where 
no muons exist, we can impose $Y_{L\mu}=Y_\mu+Y_{\nu_\mu} = 0$. 

For matter where nucleons and hyperons are the relevant hadronic degrees of freedom 
the chemical equilibrium conditions can be explicitly written as 
\begin{equation}
\begin{array}{l}
    \mu_{\Xi^-}=\mu_{\Sigma^-} = \mu_n + \mu_e - \mu_{\nu_e}, \nonumber \\
    \mu_{\Lambda} = \mu_{\Xi^0}=\mu_{\Sigma^0} = \mu_n , \nonumber \\
    \mu_{\Sigma^+} = \mu_p = \mu_n - \mu_e + \mu_{\nu_e} , \nonumber  \\  
    \mu_{\mu} - \mu_{\nu_\mu} = \mu_e - \mu_{\nu_e} \ . 
\end{array}
    \label{eq:eqc}
\end{equation}
In the case of neutrino-free matter (relevant to describe the cold and deleptonized 
neutron star) the new equilibrium conditions can be obtained by the previous 
equations simply by taking $\mu_{\nu_e} = \mu_{\nu_\mu} = 0$.

For a given value of the total baryon number density 
\begin{eqnarray}
                \rho_b = \sum_{i} \rho_{B{_i}}
\label{eq:rhot}
\end{eqnarray}
the composition of stellar matter, 
{\it i.e.,} the baryonic ($Y_{B{_i}} = \rho_{B{_i}}/\rho_b$) and leptonic  
($Y_{\ell{_i}} = \rho_{\ell{_i}}/\rho_b$)  fractions of each constituent species, 
is obtained by solving Eqs.\ (\ref{eq:charge}), (\ref{eq:lepnum}) and (\ref{eq:eqc}). 
We will refer to this status of the stellar matter as $\beta$-stable matter.    

The chemical potentials of the different particles are the fundamental ingredients 
when solving the equilibrium conditions summarized in Eq.\ (\ref{eq:eqc}). 
In the BHF approximation the chemical potentials of the baryons are taken to be equal to the value of 
the single-particle energy at the Fermi momentum, 
\begin{equation}
\begin{array}{r}
\displaystyle{
\mu_{B_i} = E_{B_i}(k_{F_{B_i}})= M_{B_i} + \frac{k_{F_{B_i}}^2}{2M_{B_i}}
+ }\nonumber \\
\displaystyle{
            U_{B_i}^N(k_{F_{B_i}})+ U_{B_i}^Y(k_{F_{B_i}}) } \ .
\end{array}
\label{eq:chempot}
\end{equation}
In order to be consistent with our calculation of the baryonic energy density 
(see the discussion above in connection to the role of nucleon TBF for the 
saturation properties of nuclear matter), in the case of nucleons,    
we replace in the chemical potentials given by Eq.\ (\ref{eq:chempot}) the nucleonic BHF 
contribution  $\mu_N^N \equiv M_N+k_{F_N}^2/2M_N + U_N^N(k_{F_N})$ 
by $\mu_N^N=\partial\varepsilon_{NN}/\partial\rho_N$. 
Here $\varepsilon_{NN}$ denotes the parametrization of the 
nucleonic energy density contribution due to Heiselberg and Hjorth-Jensen (1999) (see Eq.\ (\ref{eq:hh99})).  
For the hyperons, however, we keep the prescription of Eq.\ (\ref{eq:chempot}).  
The chemical potentials of leptons are calculated using the expressions for   
non-interacting relativistic fermions which are well known from textbooks. 

Once the composition of $\beta$-stable matter is determined we can compute 
the total energy density 
$\varepsilon = \varepsilon_b + \varepsilon_\ell$, 
the baryonic pressure using the thermodynamic 
relation 
\begin{equation}
P_b=\rho_b\frac{\partial\varepsilon_b}{\partial\rho_b}-\varepsilon_b\  \ ,
\label{eq:press2}
\end{equation}
and finally the total pressure  $P  = P_b + P_\ell$. 
Once again the leptonic contributions to the energy density and pressure are 
those of a relativistic free Fremi gas.

We note here that, although the hyperon chemical potentials are evaluated according to 
Eq.\ (\ref{eq:chempot}), and the hyperonic contribution to the nucleon chemical potentials
is keep to be $U_{N}^Y(k_{F_{N}})$, the thermodynamic relation
\begin{equation}
\varepsilon+P=\sum_{i=B,L}\rho_i\mu_i \ ,
\label{eq:relation}
\end{equation}
is fulfilled within $1\%$ at saturation density, and within $10\%$ at the central density corresponding to the 
maximum mass configuration.


\begin{figure}
\resizebox{\hsize}{!}{\includegraphics{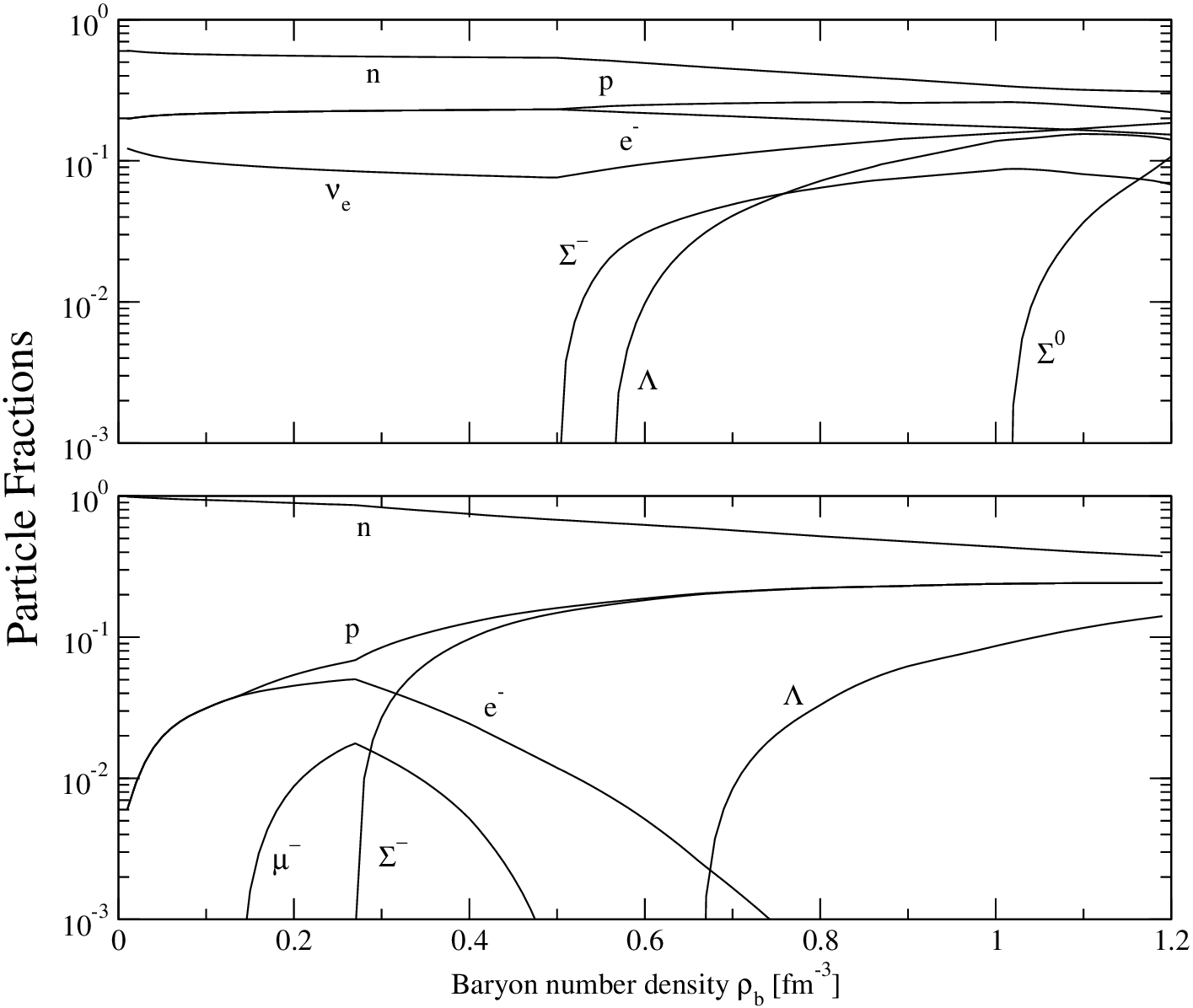}}
   \caption{Compositon of $\beta$-stable hyperonic matter as a function
of the baryon number density. Upper panel show results for the 
neutrino-trapped case (with $Y_{L_e}=0.4$ and $Y_{L_\mu}=0$), whereas
those for neutrino-free matter are reported in the lower one.}
   \label{fig:fig1}
\end{figure}

\section{Results}
\label{sec:sec3}
The composition of $\beta$-stable stellar matter calculated as described in the 
previous section, is shown in Fig.\ \ref{fig:fig1}  as a function of the total baryon number 
density. The upper panel of the figure exhibits the results for neutrino-trapped 
matter (with $Y_{Le}=0.4$ and $Y_{L\mu}=0$), whereas the lower panel shows the 
composition of neutrino-free matter.     
To begin with, let us comment our results for the composition of neutrino-free matter 
in connection to the role of the hyperon-nucleon and hyperon-hyperon interactions 
(see Ref. Vida\~na {\it et al.} 2000a, for a more detailed discussion). 
Firstly, note that although the $\Lambda$ hyperon is about $80$ MeV less massive 
than the $\Sigma^-$ one, the latter appears at a lower baryon number density. The reason is that the process
$e^-+n \rightarrow \Sigma^-+\nu_e$ removes both an energetic neutron and an energetic electron,
whereas the weak strangeness non-conserving decay of a neutron into a $\Lambda$, being neutral, removes only an
energetic neutron. Since the electron chemical potential in matter is larger than the mass difference $M_{\Sigma^-}-M_{\Lambda}$, 
the condition for the onset of the $\Sigma^-$, $\mu_n+\mu_e = \mu_{\Sigma^-}$, is fulfilled at lower densities
than the corresponding one for the appeareance of the $\Lambda$, $\mu_n=\mu_{\Lambda}$. Furthermore, as soon as the $\Sigma^-$ appears it 
becomes 
energetically more favorable for the system to keep charge neutrality with 
$\Sigma^-$  hyperons  than with leptons, therefore the lepton concentrations begin 
to fall. The onset of $\Lambda$ formation takes place at higher baryon number density as soon as 
the chemical potential of the neutron equals that of the $\Lambda$. 
No other hyperons appear at baryon number densities below $\rho_b=1.2$ fm$^{-3}$ within our 
many-body approach.

\begin{figure}
\resizebox{\hsize}{!}{\includegraphics{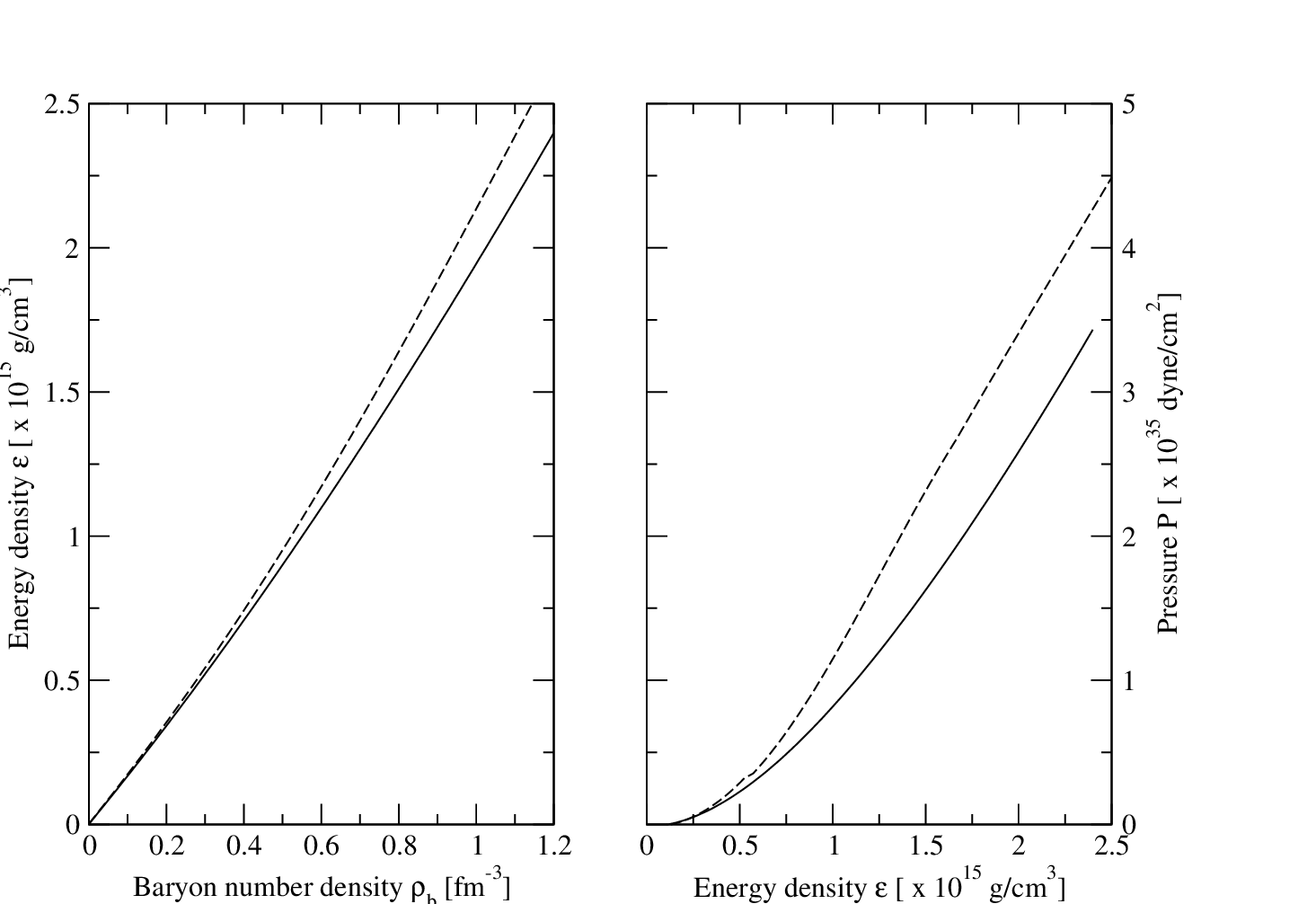}}
   \caption{Total energy density $\varepsilon$ as a function of the
baryon number density $\rho_b$ (left panel) and total pressure as a
function of 
$\varepsilon$ (right panel) for $\beta$-stable neutron star matter. Solid
lines in both panels show results for neutrino-free matter, whereas 
results for the neutrino-trapped case correspond to dashed lines.}
   \label{fig:fig2}
\end{figure}

Having in mind these results as a reference, let us now consider the effect 
of neutrino trapping. As it can be seen from the upper panel of 
Fig.\ \ref{fig:fig1} the composition of matter is significantly altered when 
neutrinos are trapped. The first thing to notice is that trapping keeps the 
electron concentration high so that matter is more proton rich in comparison with 
the case in which neutrinos have diffused out. Notice in addition that muons are not 
present, and the onset of hyperon formation is changed. The appearance of the 
$\Sigma^-$ hyperon is now governed by $\mu_{\Sigma^-}=\mu_n+\mu_e-\mu_{\nu_e}$, 
whereas in the neutrino-free case the condition to be fulfilled was 
$\mu_{\Sigma^-}=\mu_n+\mu_e$. Due to the fact that $\mu_e-\mu_{\nu_e}$ is much 
smaller than $\mu_e$, the appearance of the $\Sigma^-$ occurs at a higher baryon number density 
($\rho_b \approx 0.50$ fm$^{-3}$), and the amount of $\Sigma^-$'s is smaller. 
This, in turn, implies less $\Sigma^-n$ pairs. Since the $\Sigma^-n$ 
interaction is attractive in this model (see, {\it e.g.,} Fig.\ 7 of Ref. Vida\~na 
{\it et al.} 2000b) the chemical potential of the neutrons becomes less attractive. 
As a consequence, the $\Lambda$ and $\Sigma^0$ (which in neutrino-free matter was 
not present) hyperons appear at a lower  densities ($\rho_b \approx 0.57$ fm$^{-3}$ 
and $\rho_b \approx 1.02$ fm$^{-3}$, respectively). 
Finally, the neutrino  fraction, which initially decreases with baryon number density in order 
to keep $Y_{Le}$ constant, begins to increase as soon as $\Sigma^-$'s are present 
on the system due to the formation of this baryon through the process 
$e^- +n \rightarrow \Sigma^- + \nu_e$.

\begin{figure}
\resizebox{\hsize}{!}{\includegraphics{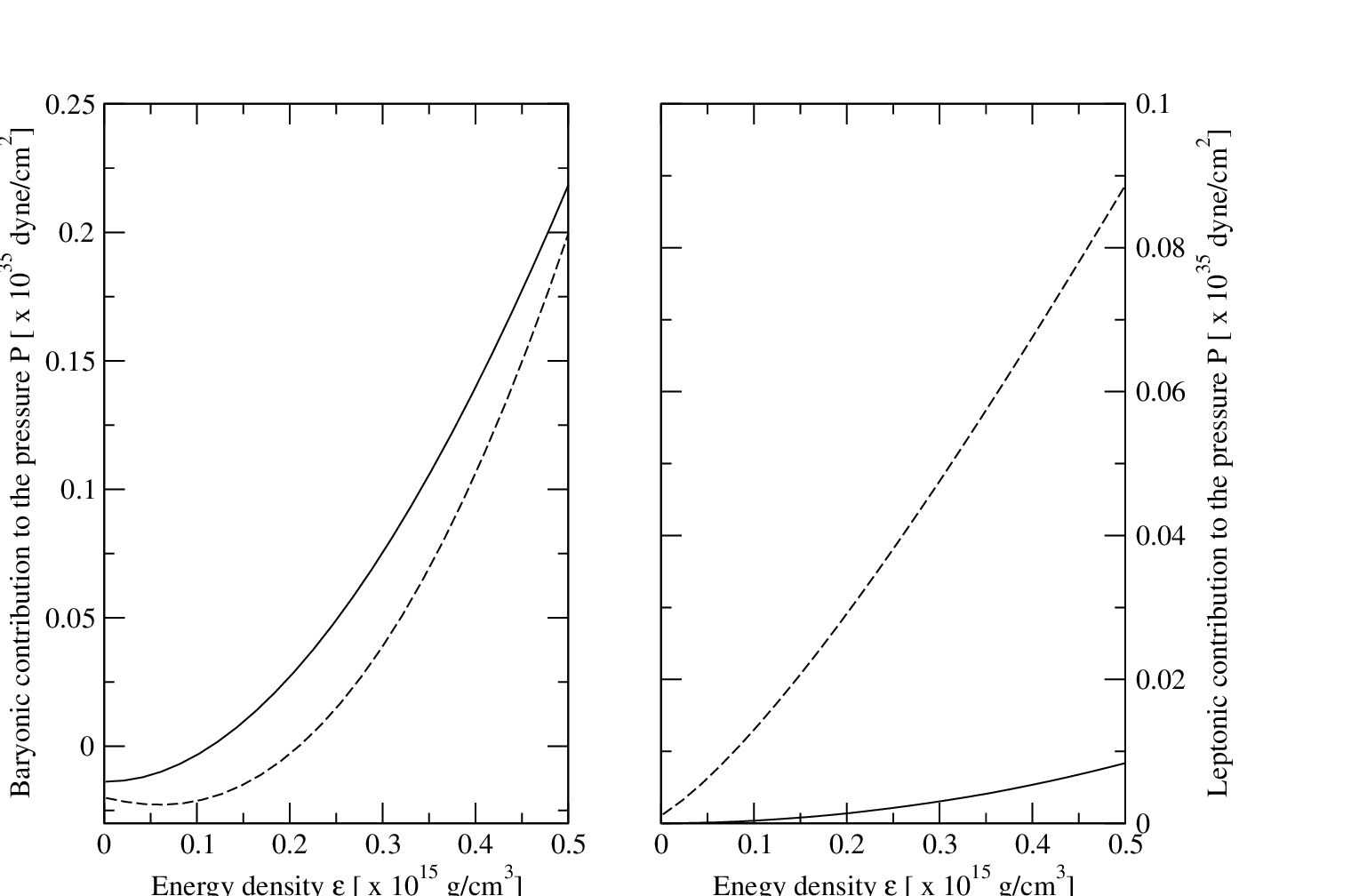}}
   \caption{Baryonic (left panel) and leptonic (right panel)
contributions to the total pressure $P$ as a function of the total energy
density 
density $\varepsilon$ for the two scenarios considered: neutrino-free
(solid lines) and neutrino-trapped (dashed lines) 
matter. The baryon number 
density corresponding to the maximum energy density plotted is $\rho_b
\sim 0.3$ fm$^{-1}$, at which the only relevant hadronic 
degrees of freedom are nucleons.}
   \label{fig:fig3}
\end{figure}

Let us now examine the effect of neutrino trapping on the EoS for $\beta$-stable 
neutron star matter. We show in Fig.\ \ref{fig:fig2} the results for the total energy density $\varepsilon$ 
versus the baryon number density (left panel) and the total pressure as a function 
of $\varepsilon$ (right panel). The dashed lines represent the results for 
neutrino-trapped matter whereas the solid lines show the result for 
neutrino-free  matter. 
As we can see the EoS for neutrino-trapped matter is stiffer than that for 
neutrino-free matter. This result is a consequence of the different composition 
of stellar matter in the two cases illustrated in  Fig.\ \ref{fig:fig1}.   
In addition, it is interesting to note that, even in those regions where nucleons 
are the only relevant baryonic degrees of freedom ({\it i.e.,} up to 
$\rho_b \sim 0.50$ fm$^{-3}$), the EoS for neutrino-trapped matter is stiffer than 
the one for neutrino-free matter. In fact, the extra leptonic pressure caused by 
neutrino-trapping is greater than the decrease in pressure of nucleons induced by 
the reduction of  the nuclear symmetry energy in the proton rich matter with 
trapped neutrinos (compare the the proton abundances in the upper and lower panel of 
Fig.\ \ref{fig:fig1} for $\rho_b < 0.50$ fm$^{-3}$) 
(Chiapparini {\it et al.} 1996, Prakash {\it et al.} 1997) . 
This can be seen in Fig.\ \ref{fig:fig3} where we plot the 
baryonic (left panel) and leptonic (right panel) 
contributions to the total pressure for neutrino-free (solid lines) and 
neutrino-trapped (dashed lines) matter. 

Finally, let us consider the effect of neutrino trapping on the properties of 
neutron stars. 
To this end, we have solved the well known Tolman--Oppenheimer--Volkov 
equations for the structure of non-rotating stellar configurations in general 
relativity. 
To describe the stellar crust we used the equations of state by 
Feynman--Metropolis--Teller (Feynman {\it et al.} 1949), Baym--Pethick--Sutherland 
(Baym {\it et al.} 1971) and Negele-Vautherin (1973). 
In Fig.\ \ref{fig:fig4} we show the resulting stellar equilibrium sequences. 
In the left panel we plot the gravitational mass $M_G$ in units of the solar mass 
($M_\odot = 1.989 \times 10^{33}$~g) as a function of the central energy density,  
while in the right panel $M_G$ is plotted  as a function of the stellar radius $R$. 
Dashed (solid) lines represent the results for neutrino-trapped (neutrino-free) 
matter. The properties of the maximum mass configurations are summarized in Table \ref{tab:tab1}. 
The EoS for neutrino-free matter calculated within the present microscopic approach (Baldo {\it et al.} 2000,
Vida\~na {\it et al.} 2000a) gives a maximum mass below $\sim 1.44 M_\odot$, in conflict with measured neutron 
star masses. This means that our EoS with hyperons needs to be stiffer. Within a microscopic approach, as the one
used in this work, one should try to trace the origin of this problem back to the underlying hyperon-nucleon and 
hyperon-hyperon two-body interaction, or to the possible repulsive three-body forces involving one or more hyperons
(i.e., YNN, YYN or YYY), not included in this work and similar studies. Unfortunally, the YY two-body interaction
is not well constrained at present due to the scarce amount of experimental data, and although active research is 
devoted to the constructuion of three-body forces between nucleons and hyperons they are not yet available.
On the other hand the neglect of the hyperonic degrees of freedom on the dense matter EoS
always leads to an unrealistic overstimate of the stellar maximum mass.

\begin{figure}
\resizebox{\hsize}{!}{\includegraphics{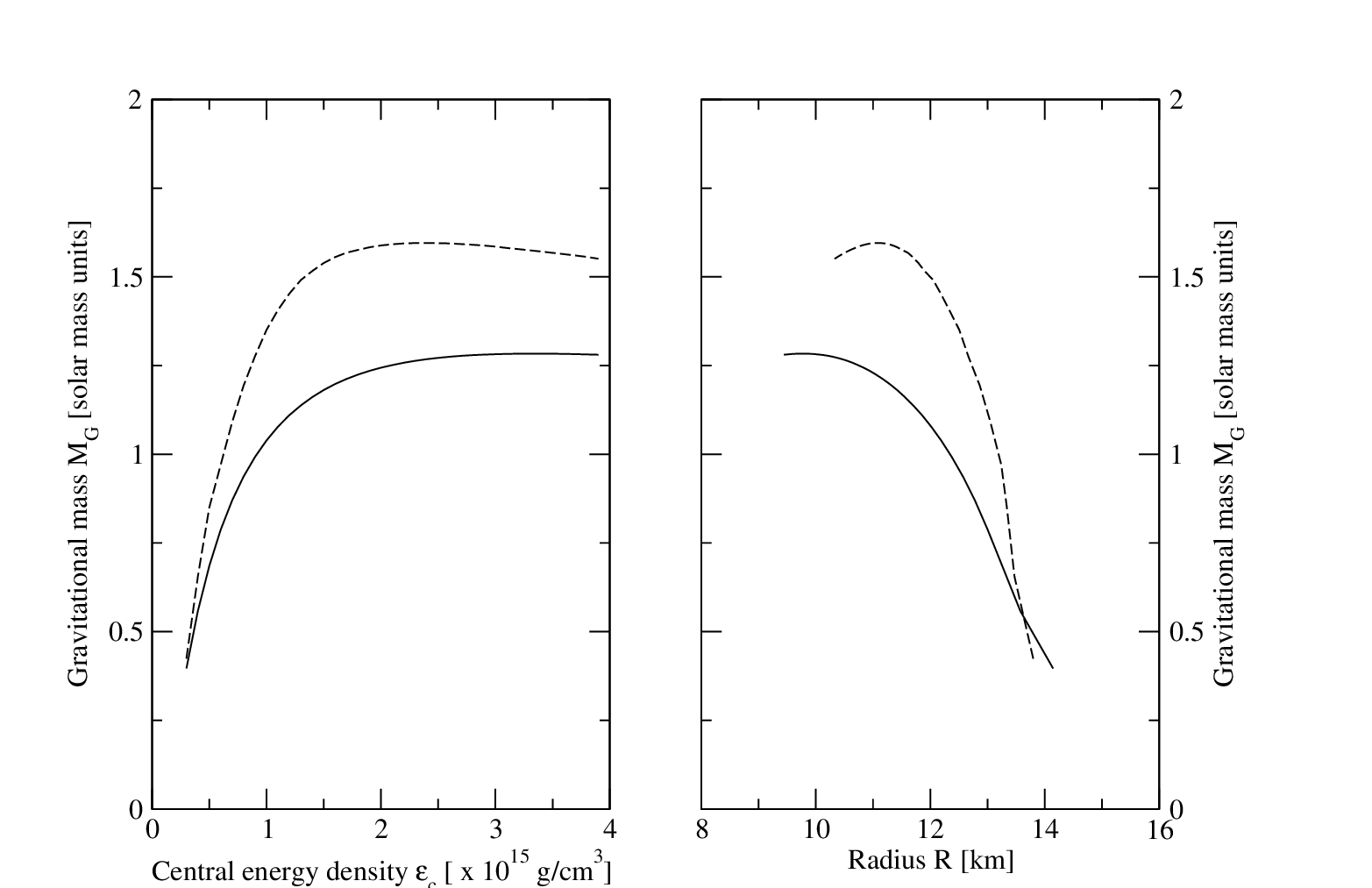}}
   \caption{Gravitational mass as a function of the central energy density (left panel) and radius (right panel) of the star
for the two scenarios considered: neutrino-free (solid lines) and neutrino-trapped matter (dashed lines).}
   \label{fig:fig4}
\end{figure}


\begin{table}
\centering
\caption{Neutron star properties of the maximum mass configuration for
the two scenarios considered: neutrino-trapped and neutrino-free 
matter. $\varepsilon_c$ denotes the central energy density, $\rho_{b_c}$
the corresponding central baryon number density, $M_G$ the gravitational 
mass, $M_B$ the baryonic mass, $R$ the radius of the star, $R_Y$ the
radius of the hyperonic core, and $\Delta$$R_{crust}$ the thickness of
the 
star crust.}
\bigskip
\bigskip
\begin{tabular}{c| cccc}
\hline
\hline
\cr
Scenario & $\varepsilon_c$ & $\rho_{b_c}$ & $M_G$ & $M_B$ \cr
    &[$\times 10^{15}$ g/cm$^3$]  & [fm$^{-3}$] & [$M_\odot$] &
[$M_\odot$] \cr
\cr
\hline
\cr
Trapped & $2.30$ & $1.066$ & $1.595$ & $1.724$ \cr
\cr
Free    & $3.19$ & $1.537$ & $1.283$ & $1.406$ \cr
\cr
\hline
\hline
\end{tabular}

\vskip0.2cm
\begin{tabular}{ccc}
\hline
\hline
\cr
$R$ & $R_Y$ & $\Delta R_{crust}$ \cr
[km]  & [km] & [km] \cr
\cr
\hline
\cr
$11.14$ & $6.32$ & $0.66$ \cr
\cr
$9.86$ & $7.60$ & $0.70$ \cr
\cr
\hline
\hline
\end{tabular}
\label{tab:tab1}
\end{table} 

In agreement with previous studies 
we find that the maximum mass supported by neutrino-trapped EoS is larger than 
the corresponding one supported by neutrino-free matter EoS. 
The overall effect of neutrino trapping on the maximum mass configuration 
is opposite in the case of matter in which the only baryonic degrees of freedom 
considered are nucleons (Bombaci 1996, Prakash {\it et al.} 1997). 
In the latter case, the lost of leptonic pressure when neutrinos 
are diffused out of the star is smaller than the gain in baryonic pressure arising 
from the nuclear symmetry energy due to the decrease in the number of protons. 
As a consequence, in nucleonic $\beta$-stable matter, 
the maximum mass supported by neutrino-free matter is larger than the 
corresponding one supported by neutrino-trapped matter, as it is shown by our present 
results reported in the right panel of Fig.\ \ref{fig:fig5}.   

A very important implication of neutrino trapping in dense matter with hyperons  
is the possibility of having metastable neutron stars and a delayed formation 
of a ``low-mass'' ($M = 1$ -- 2 $M_\odot$) black hole. 
This is illustrated in Fig.\ \ref{fig:fig5} where we show the gravitational mass 
of the star as a function of its baryonic mass $M_B$, which is taken as  
the total number of baryons in the star times the average nucleon mass.    
If hyperons are present (left panel), then deleptonization lowers the range of 
gravitational masses that can be supported by the EoS from about $1.59 M_\odot$ 
to about $1.28 M_\odot$ (see dotted horizontal lines in the figure). 
Since most of the matter accretion on the forming neutron star happens in the 
very early stages after birth ($ t < 1$~s), with a good approximation, 
the neutron star baryonic mass stays constant during the evolution from the 
initial protoneutron star configuration to the final neutrino-free configuration. 
Then, within our EoS model, protoneutron stars which at birth have 
a gravitational mass between  1.28 -- 1.59 $M_\odot$ 
(a baryonic mass between  1.40 -- 1.72 $M_\odot$) will be stabilized by neutrino 
trapping effects long enough to carry out nucleosynthesis accompayning a 
type-II supernova explosion.   
After neutrinos leave the star, the EoS is softened and it can not support anymore 
the star against its own gravity. Thus the newborn neutron star collapses to 
a black hole (Keil and Janka 1994, Bombaci 1996, Prakash {\it et al.} 1997). 
A similar qualitative behaviour is expected also in the case in which dense matter 
contains a Bose--Einstein condensate of negative kaons (Brown \& Bethe 1994, 
Prakash {\it et al.} 1997).   
On the other hand, if only nucleons are considered to be the relevant baryonic 
degrees of freedom (right panel), no metastability occurs 
and a black hole is unlikely to be formed during the deleptonization since 
the gravitational mass increases during this stage which happens at 
constant baryonic mass. If a black hole were to form from a star with only nucleons, 
it is much more likely to form during the post-bounce accretion stage. 

\begin{figure}
\resizebox{\hsize}{!}{\includegraphics{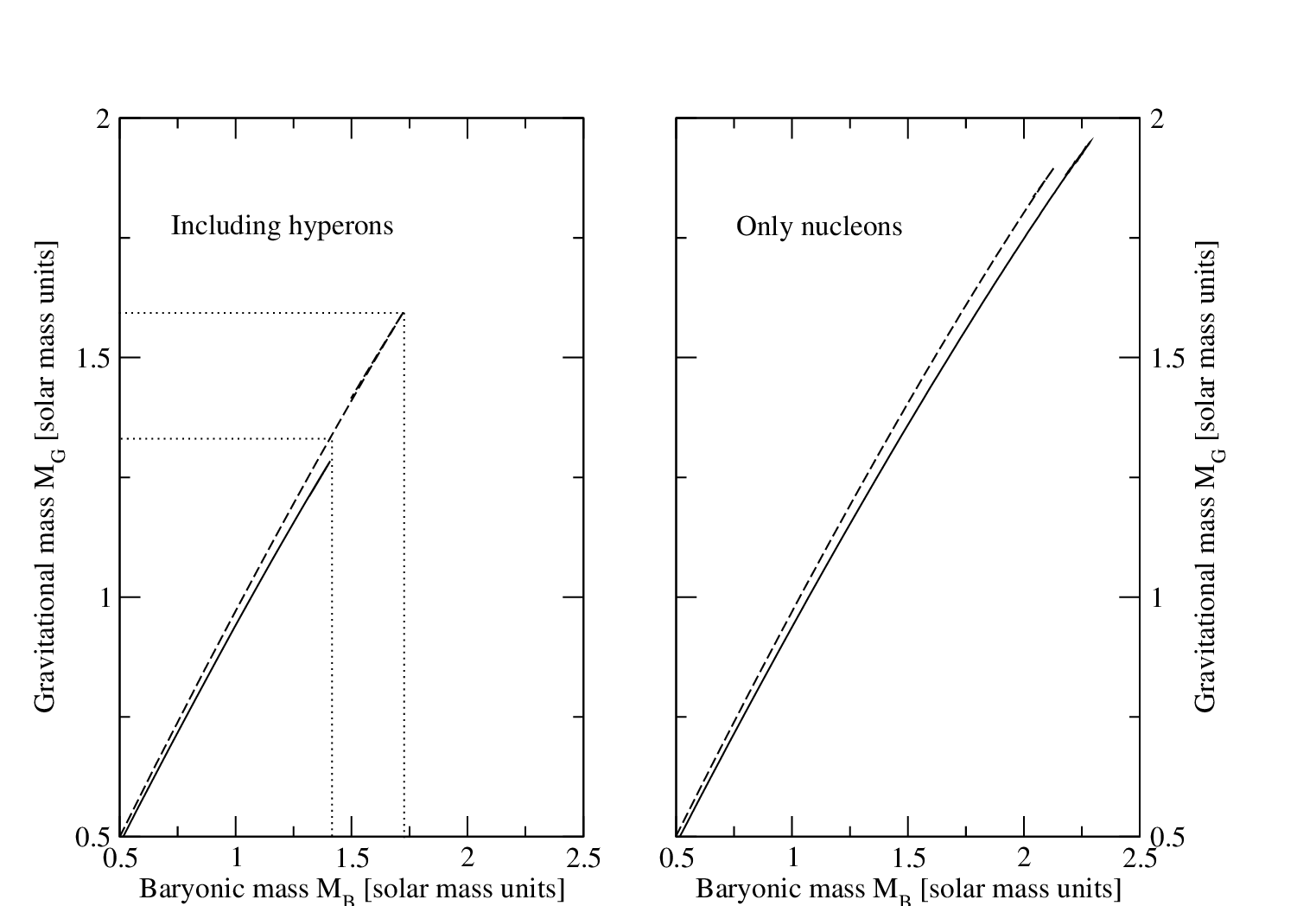}}
   \caption{Gravitational mass as a function of the baryonic mass
for the two scenarios considered: neutrino-free (solid lines) and neutrino-trapped matter (dashed lines).
Left panel shows results for matter containing nucleons and hyperons as baryonic degrees of freedom, whereas results containing only nucleons are 
reported
on the right one. Dotted lines on the left panel show the window of metastability in the gravitational and baryonic masses.}
   \label{fig:fig5}
\end{figure}

\begin{figure}
\resizebox{\hsize}{!}{\includegraphics{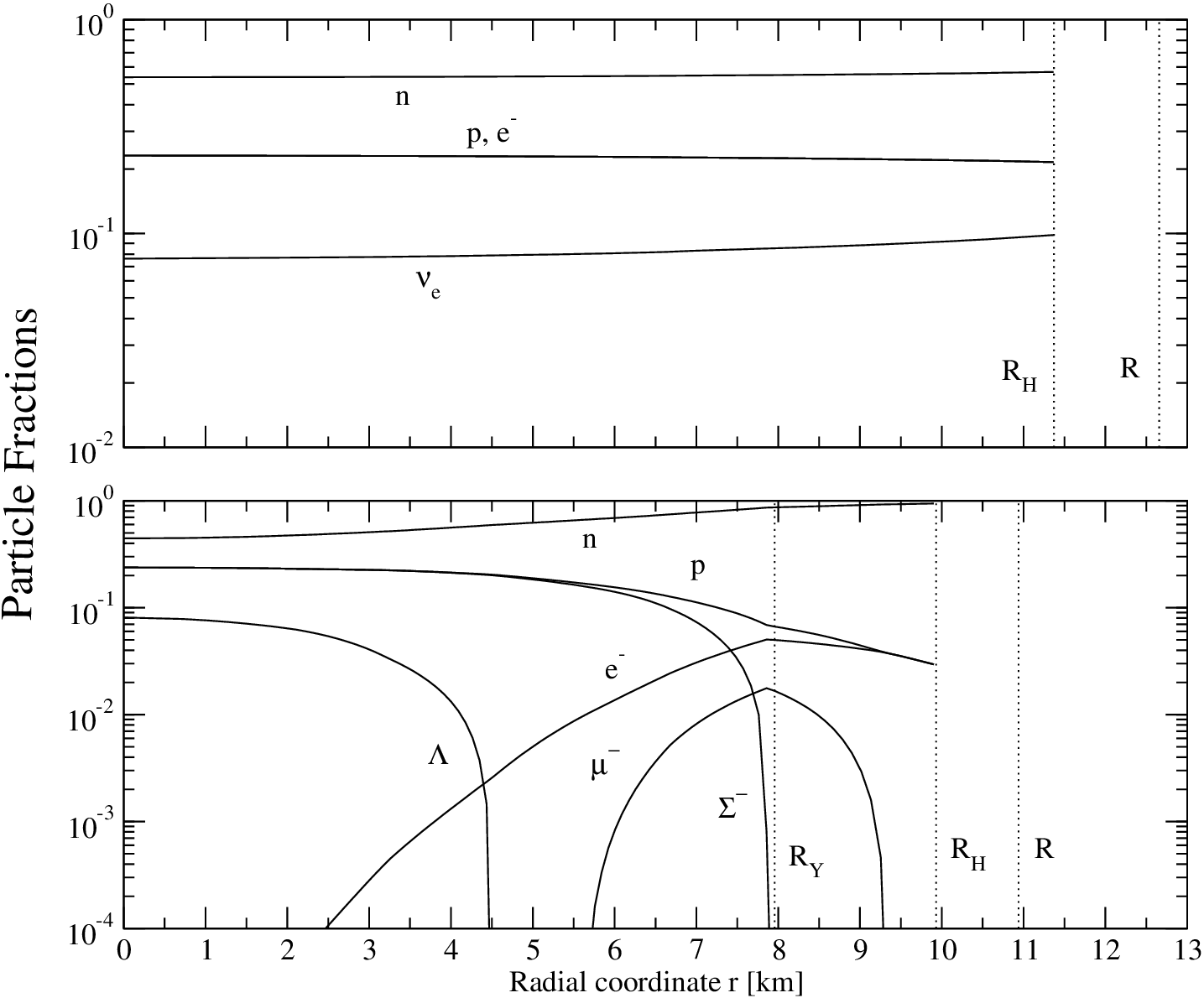}}
   \caption{Internal composition as a function of the radial coordinate of a hyperon star of constant baryonic mass ($M_B=1.34 M_{\odot}$) as it 
evolves from the initial neutrino-trapped (upper panel) to the final neutrino-free (lower panel) configuration. Symbol $R$ indicates the radius 
of the star, $R_H$ the end of the hadronic core and the beginning of the crust and $R_Y$ the end of the hyperonic core.}
   \label{fig:fig6}
\end{figure}

To end this section, we show in Fig.\ \ref{fig:fig6} the differences of 
the internal composition as a function of the radial coordinate of a protoneutron 
star (upper panel) and the corresponding deleptonized neutron star (lower panel) 
for a constant value, $M_B = 1.34 M_\odot$, of the stellar baryonic mass.   
The central energy density of the protoneutron star is not high enough to allow for the 
presence of hyperons and only nucleons, electrons and neutrinos are present 
in the stellar core. This star has a gravitational mass $M_G = 1.28 M_\odot$.  
Nevertheless, as soon as neutrinos diffuse out of the star, pressure decreases, 
gravity compresses matter, energy density increases and hyperons appear in the star interior. 
The gravitational mass of the final neutrino-free star is $M_G = 1.23 M_\odot$. 
The difference between the initial and final gravitational masses corresponds to 
the energy which is carried out by neutrinos when they escape from the star.  
In the present case ({\it i.e.,} assuming $M_B = 1.34 M_\odot$) this energy is 
about $9 \times 10^{52}$~erg. In addition, due to the increase of the central energy density,   
the stellar radius decreases.


\section{Summary and Conclusions}
\label{sec:sec4}

In this paper we have investigated within the framework of the 
Brueckner--Hartree-Fock approximation the effects  of neutrino trapping on the 
properties of $\beta$-stable neutron star matter including nucleonic and hyperonic 
degrees of freedom. 

We have found that the presence of neutrinos changes 
significantly the compositon of matter with respect to the neutrino-free case: 
matter becomes more proton rich, muons are not present, and the appearance of 
hyperons is moved to higher densities. In additon, the number of strange particles 
is on average smaller and the EoS stiffer in comparison with the neutrino-free case.

We have found that the value of the maximun mass of hyperon stars decreases 
as soon as neutrinos diffuse out of the star, contrary to what happens when the only baryonic 
degrees of freedom considered are nucleons.

Using the microscopic EoS developed in the present work we have found 
that stars having at birth a gravitational mass between  1.28 -- 1.59 $M_\odot$  
are metastable, in other words these stellar configurations remain 
only stable for several seconds (the neutrino trapping time), collapsing afterwards   
into low-mass black holes. 


\end{document}